\documentclass[a4paper,11pt]{llncs}

\usepackage{amssymb,amsmath,
graphics,color}
\usepackage{epsfig}
\usepackage{url}

\newenvironment{proofof}{\noindent\textit{\textbf{Proof of}}}{{\hfill $\Box$}}

\DeclareMathOperator{\LinPath}{LinPath}

\begin{document}

\title{\textbf{Abstract Milling with Turn Costs
\thanks{Research initiated at the 6th McGill - INRIA Barbados
Workshop on Computational Geometry in Computer Graphics, 2007.}
}}

\author{
M. Fellows\inst{1}
\thanks{Research supported by the Australian Research Council through the ARC Centre of Excellence in Bioinformatics and 
Discovery Project DP0773331.}
\and 
P. Giannopoulos\inst{2}\thanks{Research supported by the German Science Foundation (DFG) under grant Kn 591/3-1.}
\and 
C. Knauer\inst{2}
\and 
Christophe Paul\inst{3}
\thanks{Research conducted while the author was on sabbatical at McGill University,
School of Computer Science, Canada.  Supported by the project
ANR-06-BLAN-0148.}
\and 
F.~Rosamond\inst{1}
\thanks{Research supported by the Australian Research Council.}
\and 
S. Whitesides\inst{4}
\thanks{Research supported by NSERC and FQRNT.}
\and 
N. Yu\inst{4}
}

\institute{PCRU, Office of DVC(Research), University of Newcastle, Australia 
\and 
Institut f\"ur
Informatik, Freie Universit\"at, Berlin, Germany 
\and
CNRS - LIRMM, Montpellier, France
\and
McGill University, School of Computer Science, Canada.
}

\maketitle

\begin{abstract}
The {\sc Abstract Milling} problem is a natural and quite general graph-theoretic model
for geometric milling problems. Given a graph, one asks
for a walk that covers all its vertices with a minimum number of {\it turns}, as
specified in the graph model by a 0/1 turncost function $f_{x}$ at each vertex $x$ giving, 
for each ordered pair of edges $(e,f)$ incident at $x$, the {\it turn cost} at $x$ of a 
walk that enters the vertex
on edge $e$ and departs on edge $f$.
We describe an initial study of the parameterized complexity of the 
problem. 
Our main positive result shows that 
{\sc Abstract Milling}, parameterized by: number of turns, treewidth
and maximum degree,
is fixed-parameter tractable,
We also  
show that {\sc Abstract Milling}  
parameterized by (only) the number of turns and the pathwidth,
is hard for $W[1]$ --- one of the few parameterized
intractability results for bounded pathwidth.   
\end{abstract}

\section{Introduction}


We consider the following problem:

\noindent
{\sc Abstract Milling} \\
{\it Instance:} A simple graph $G=(V,E)$ and for each vertex $x$, a 
{\it turncost function} $f_x$ indicating whether a {\it turn} is required, with $f_{x}: E(x) \times E(x) \rightarrow 
\{0,1\}$, where $E(x)$ is the set of edges incident on $x$.\\
{\it Question:} Is there a walk making at most $k$ turns that visits every vertex of $G$? 

The {\sc Grid Milling} problem restricts the
input to \emph{grid graphs}, that is, rectilinearly plane-embedded graphs 
that are subgraphs of the integral grid, with the obvious and natural turn cost functions.

\noindent
{\bf Results.}
Our basic starting point is an FPT algorithm for {\sc Grid Milling}, parameterized by the
numbers of turns.
Generalizing this, we give an FPT result for {\sc Abstract Milling}, 
parameterized by $(k,t,d)$, where
$k$ is the number of turns, $t$ is the tree-width of the input graph $G$,
and $d$ is the maximum degree of $G$.  
Next, we explore whether this positive result can be further strengthened.  However: 
the {\sc Abstract Milling} problem is $W[1]$-hard when parameterized by $(k,p)$, where $k$ is the number of turns
and $p$ is the path-width of $G$ (and therefore also when parameterized by $(k,t)$). (This hardness result is 
actually shown for a restricted version of the problem, called {\sc Discrete Milling}, see below).

\noindent
{\bf Significance.}
The {\sc Abstract Milling} problem is motivated by (and generalizes) similar graph-theoretic models of geometric milling
problems introduced by Arkin et al. \cite{ABD05}, and our results are concretely interesting as the first
investigation of the parameterized complexity of these problems.
Recently, there has been increasing attention to the complexity of
``highly structured graph problems'' parameterized by treewidth or pathwidth.  
Our negative result provides
one of the few natural problems known to be $W[1]$-hard when the parameter
includes a pathwidth bound.

\noindent
{\bf Method and Practicality.}  Our FPT results are based on a general form of Courcelle's Theorem about
decidability of MSO properties of relational structures of bounded treewidth \cite{Cou90}, and are not
practical, although, of course leaving open the possibility that these FPT classifications may be
superseded by practical FPT algorithms, as has often occurred in the study of fixed parameter algorithms
\cite{Nie06}.

\noindent
{\bf Previous and Related Work.}
\emph{Geometric milling} is a common problem in manufacturing
applications such as numerically controlled  pocket machining and
automatic tool path generation; see Held~\cite{H91} for a survey.
The {\sc Discrete Milling} problem 
introduced by Arkin et al.~\cite{ABD05} 
uses a graph model to study 
milling problems with turn costs and other
constraints.  A solution
path must visit a set of vertices 
that are connected by edges representing the
different directions (``channels'') that the ``cutter'' can take.
In the model introduced by Arkin et al., incident edges to a vertex $x$ are {\it paired}
in the cost function $f_x$ in the sense that for each incident edge $e$ there is at most one incident
edge $f$ such that $f_{x}(e,f)=0$, and symmetric: if $f_{x}(e,f)=0$ then $f_{x}(f,e) =0$.
We consider here the more general {\sc Abstract Milling} problem, that allows an arbitrary
0/1 turncost function at each vertex.

Arkin et al.~\cite{ABD05} showed that {\sc Discrete 
Milling} is NP-hard (even for grid graphs) and described a constant-factor
approximation algorithm for minimizing
the number of turns in a solution walk. 
They also described a PTAS
for the case where the cost is a linear combination of the length of the walk and
the number of turns.  

\noindent
{\bf Definitions and Preliminaries.}
We will assume that the basic ideas of parameterized complexity theory and bounded tree-width algorithmics
up through the basic form of Courcelle's Theorem and monadic second-order logic (MSO) are known to the reader; 
some basic definitions are provided in Section \ref{sec:MSO} and in Appendix B.  Details of routine deployments
of MSO in the proofs of our theorems (that can be laborious in full formality) are relegated to Appendix A due to space limitations.
For background on these topics, see \cite{DF99,FG06}.

For a graph $G$, let $\text{tw}(G)$ be its treewidth.
We assume that all graphs $G$ are simple (no loops or multiple edges).
A \emph{walk} $W=[x_0,\dots, x_l]$ on a graph $G=(V,E)$ is a
sequence of vertices such that every pair $x_{i}$, $x_{i+1}$ of consecutive vertices
of the sequence are adjacent (we use $x_{i}x_{i+1}$ to refer to the edge between them).
The {\it turn cost} of a
walk $W$
is defined as 
$$\text{tc}(W)=\displaystyle\sum_{i=1}^{l-1} f_{x_i}(x_{i-1}x_i,x_ix_{i+1})\, .$$
A walk that visits every vertex of a graph is termed a {\it covering walk}.
Note that in {\sc Abstract Milling} a solution covering walk may visit a vertex 
{\it many times}.

\medskip
The paper is organized as follows: Section \ref{sec:MSO} reviews the basic ideas of MSO (and its extensions) 
that we need for proving our FPT results for {\sc Grid Milling} (Section \ref{sec:grid-milling})  
and {\sc Abstract Milling} (Section \ref{sec:abstract-milling}). The intractability result for {\sc Discrete Milling} 
is shown in Section \ref{sec:discrete-milling-hardness}. We conclude with some open problems.  

\section{Monadic Second Order Logic }
\label{sec:MSO}

The usual MSO logic of graphs
can be extended to digraphs and mixed graphs (some edges are oriented and some are not),
where the vertices and edges (or arcs) have a fixed number of types. 
This was proved in full generality
first by Courcelle \cite{Cou90}, and is exposited well in \cite{FG06} in terms of relational structures of bounded
treewidth.
We refer to such a mixed graph with a fixed number of types of edges,
arcs and vertices as an {\it annotated} graph.  The treewidth of an annotated graph is the treewidth of the
underlying undirected graph.  

In this paper, we will associate to each input graph an annotated
graph, in such a way that the property of being a yes-instance of the problem under consideration can be
expressed as an MSO property of the associated annotated graph. 
The formal means that MSO logic (as we will use it here) provides us with, and Courcelle's Theorem 
can be found in Appendix B.

\section{Grid Milling is Fixed-Parameter Tractable}
\label{sec:grid-milling}

We prove here our starting point: that {\sc Grid Milling} is FPT for parameter $k$,
the number of turns.  We first argue that instances
with large tree-width are no-instances.  Then we show how to express 
the problem in 
(extended) MSO for an annotated graph ${\cal M}(G)$
that we associate to a {\sc Grid Milling} instance $G$.
That is, we describe an MSO formula $\phi$, such that 
the property expressed by $\phi$ holds for
${\cal M}(G)$
if and only if $G$ is a yes-instance for the {\sc Grid Milling} problem.

\begin{lemma} \label{lem:tw-grid}
Let $G=(V,E)$ be a connected grid graph with $\text{tw}(G) > 6k-5$.
Then $G$ does not contain a $(k-2)$-turn covering walk.
\end{lemma}
\begin{proof}
We show that $G$ contains $k$ vertices that have pairwise 
different $x$- and $y$-coordinates. Then, any covering walk needs to take 
at least one turn between any two such vertices, and thus it needs at least 
$k-1$ turns in total. 

Since $G$ is planar and $\text{tw}(G) > 6k-5$, by the Excluded Grid Theorem for planar graphs (c.f. \cite{FG06}), 
it has a $(k\times k)$-grid $H$ as a minor. $H$ contains $k/2$ vertex-disjoint nested cycles.
Since taking minors can destroy or merge cycles but not create 
completely new ones, in the ``pre-images'' (under the operation taking minor) 
of these cycles there must be $k/2$ vertex-disjoint subgraphs in $G$, 
each containing a cycle. Thus, $G$ contains a set $\mathcal{C}$ of $k/2$ vertex-disjoint cycles, which must 
also be nested.
Consider a straight line $L$ of unit slope that intersects the innermost 
cycle of $\mathcal{C}$ at two vertices (grid points). $L$ must also 
intersect every other cycle at at least two vertices. 
This produces a set of at least 
$k$ intersection vertices in $G$ with the claimed property.
\qed
\end{proof}

We associate to the grid graph $G$ a (closely related) annotated graph
${\cal M}(G)$: we simply regard the horizontal edges as being of one type, and the vertical
edges as being of a second type.  Equivalently, we can think of $G$ as presented to us
with a partition of the edge set: $G = (V,E_{h},E_{v})$.  

Then, intuitively, $G$ has a
$k$-covering walk if and only if there exist
a start vertex $v_0$, turn vertices $v_1$,\ldots ,$v_k$, an end vertex $v_{k+1}$, 
and sets of vertices $S_0$,\ldots ,$S_k$, such that:
 
\noindent(i) the graph induced by $S_i$, $i=0,\ldots ,k$, is a monochromatic path, 
i.e. a path whose edges are all either in $E_h$ or in $E_v$,

\noindent(ii) the path induced by $S_i$ starts at $v_i$ and ends at $v_{i+1}$, and 

\noindent(iii) $V=\cup S_i$, i.e. all vertices of $G$ are covered. 

This is straightforwardly formalized in MSO (see Appendix A).

\begin{lemma} \label{lem:msol-grid}
Let $G=(V,E_h,E_v)$ be a grid graph.  The property of having a
$k$-covering walk on $G$ is expressible in MSO.
\end{lemma}

Trivially, the model graph ${\cal M}(G)$ has treewidth bounded as a function
of the treewidth of the original graph $G$.
From Lemmata~\ref{lem:tw-grid},~\ref{lem:msol-grid} and Courcelle's Theorem we have:

\begin{theorem}\label{th:grid}
\textsc{Grid Milling} is FPT with respect to k (number of turns). 
\end{theorem}

\section{Extending Tractability}
\label{sec:abstract-milling}

What makes the {\sc Grid Milling} problem FPT?  A few properties of
grid graphs might lead us to tractable generalizations: 
(i) Yes-instances must have bounded treewidth, (ii) vertices in grid graphs have bounded degree, and 
(iii) the turn-cost function is pairing and symmetric, as in the more general
{\sc Discrete Milling} problem.

We are naturally led to three questions, by relaxing these conditions: \\
$\bullet$  What is the complexity of {\sc Abstract Milling} parameterized by $(k,t,d)$, where
$k$ is the number of turns, $t$ is a treewidth bound, and $d$ is a bounded on maximum degree? \\
$\bullet$  What is the complexity of {\sc Discrete Milling} parameterized by $(k,t)$? \\
$\bullet$  What is the complexity of {\sc Discrete Milling} parameterized by $(k,d)$?

In the remainder of this paper, we answer the first two.  The third question remains open.

\begin{theorem} \label{th:ktd}
\textsc{Abstract Milling} is FPT for parameter  $(k,t,d)$,
where $k$ is the number of turns, $t$ the tree-width of the graph $G$
and $d$ is the maximum degree of $G$.
\end{theorem}

\begin{proof}
We describe how an instance of the {\sc Abstract Milling} problem, consisting of $G$ and the turncost functions, can be represented by an annotated digraph
$\mathcal{M}(G)$, that allows us to use MSO logic to express a property 
that corresponds to the question that the {\sc Abstract Milling} problem asks. 
The proof therefore consists of three parts: (1) a description of $\mathcal{M}(G)$, (2) some crucial arguments
that establish necessary and sufficient criteria regarding $\mathcal{M}(G)$, for the instance of
{\sc Abstract Milling} to be a yes-instance, and (3) the expression of these criteria in MSO logic.  

Let $G=(V,E)$ be the graph of the {\sc Abstract Milling} instance.  
The vertex set of the digraph $\mathcal{M}(G)$ is $\mathcal{V}= \mathcal{V}_{1} \cup \mathcal{V}_{2}$ where
$$\mathcal{V}_{1} = \{ l[v]: v \in V \} \mbox{~~and~~} \\
\mathcal{V}_{2} = \{ t[e]: e \in E \} \cup \{t'[e]: e \in E \} $$

Intuitively (see Figure 1), we ``keep a copy'' of the vertex set $V$ of $G$, mnemonically ``$l[v]$'' for $v$, 
as a vertex location we might be during a solution walk in $G$.  Each edge $e$ of $G$ is replaced by two vertices
$t[e]$ and $t'[e]$ that represent a ``state'' in a solution walk: traversing $e$ in one direction or the
other.  In order to distinguish the directions, consider that the vertex set $V$ of $G$ is linearly ordered.
Let $e=uv \in E$ with $u < v$ in the ordering.  Our convention will be that $t[e]$ represents a traversal of $e$
from $u$ to $v$, and that $t'[e]$ represents a traversal of $e$ in the direction from $v$ to $u$.  Thus each
edge $e$ of $G$ is represented by two vertices in $\mathcal{M}(G)$.

In describing arcs of the digraph model $\mathcal{M}(G)$ we will use the notation $x \cdot y$ to denote an arc
from $x$ to $y$.
The arc set of the digraph $\mathcal{M}(G)$ is 
$$\mathcal{A} = \mathcal{A}_{1} \cup \mathcal{A}_{2} \cup \mathcal{A}_{3} \cup \mathcal{A}_{4} \cup \mathcal{A}_{5}$$
where
\begin{eqnarray*}
\mathcal{A}_{1} & = & \{t[e] \cdot l[v]: e=\{u,v\} \in E \mbox{~with~} u<v \} \\
\mathcal{A}_{2} & = & \{l[u] \cdot t[e]: e=\{u,v\} \in E \mbox{~with~} u<v \} \\
\mathcal{A}_{3} & = & \{t'[e] \cdot l[u]: e=\{u,v\} \in E \mbox{~with~} u<v \} \\
\mathcal{A}_{4} & = & \{l[v] \cdot t'[e]: e=\{u,v\} \in E \mbox{~with~} u<v \} 
\end{eqnarray*}

Let $\mathcal{A}'$ denote the union of these four sets of arcs.
Intuitively, the arcs of $\mathcal{A}'$ just ``attach'' the vertices of the digraph 
that represent edges in $G$ to the vertices of the digraph that
represent the endpoints of the edge, so that the orientations of the arcs are compatible with the interpretation of a
vertex of $\mathcal{V}_{2}$ as representing, say, a traversal of the edge $uv$ in the direction from $u$ to $v$;
the vertex therefore has an arc {\it to it} from $l[u]$ and an arc {\it from it} to $l[v]$.  An inspection of Figure 1
will help to clarify.

The arc set $\mathcal{A}_{5}$ is more complicated to write down formally.  Its mission is to record the possibilities
for cost-free passages through vertices of a solution walk in $G$.  Suppose $a$ is an arc in $\mathcal{A}'$.
Then $a$ is {\it to} or {\it from} either a $t[e]$ vertex, for some $e$, or a $t'[e]$ vertex for
some $e$.  Let $\epsilon(a)$ be defined to be this edge $e$ of $G$.  This is well-defined.  We can then define
\begin{eqnarray*}
\mathcal{A}_{5} = \{x \cdot y: x,y \in \mathcal{V}_{2},~ \exists z=l[v] \in \mathcal{V}_{1} 
\mbox{~and~} \exists a,b \in {\mathcal{A}' ~~~~~~~~~~~} \\
\mbox{~~with~~} a=x \cdot z \mbox{~~and~~} b=z \cdot y \mbox{~~and~~} f_{v}(\epsilon(a),\epsilon(b))=0 \} 
\end{eqnarray*}

We regard $\mathcal{M}(G)$ as an annotated digraph, in the sense that there are two kinds of vertices,
those of $\mathcal{v}_{1}$ and those of $\mathcal{v}_{2}$, and two kinds of arcs, those of
$\mathcal{A}'$ and those of $\mathcal{A}_{5}$.  

\begin{figure}[htb!]
\centering \resizebox{!}{1.8in} {\includegraphics{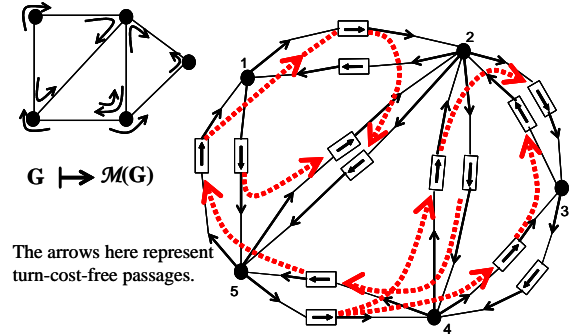}}
\caption{The arrows drawn near the $G$ vertices represent the turncost functions, indicating the
zero-cost possibilities.  These become arcs in the digraph $\mathcal{M}(G)$.}
\end{figure}

The rest of the proof will show that that the question of whether $G$ admits a covering walk making at most $k$
turns is represented by a property of the annotated digraph ${\cal M}(G)$ that can be expressed in MSO logic.  However,
before proceeding to that, it is important to verify that if the treewidth of $G$ is bounded by $t$, then the
treewidth of $\mathcal{M}(G)$ is bounded by a function of the parameter.  This depends crucially on the fact that 
the maximum degree of $G$ is part of our compound parameterization.

Suppose $\mathcal{T}(G)$ is a tree-decomposition of $G$ of width at most $t$.  We can describe a bounded width
tree-decomposition $\mathcal{T}'$ of $\mathcal{M}(G)$ as follows.
Without confusion, henceforth in this argument consider $\mathcal{M}(G)$ as an undirected graph by forgetting all
arc orientations.  
Use the same bag-indexing tree for $\mathcal{T}'$ as for $\mathcal{T}(G)$.
Suppose $B \subseteq V$ is a bag of $\mathcal{T}(G)$.  Replace $B$ with the union of the closed neighborhoods
of the vertices  of $\mathcal{V}_{1}$ corresponding to the vertices of $B$, in $\mathcal{M}(G)$.
It is easy to check that all the axioms for a tree-decomposition hold, and that the treewidth of 
$\mathcal{M}(G)$ is therefore bounded by $2dt$.

In a digraph $D=(V,A)$, by a {\it purposeful set of arcs} $(S,s,t)$ 
we refer to a set of arcs $S \subseteq A$, together with
two distinguished vertices $s,t \in V$.  
We say that a purposeful set of arcs $(S,s,t)$ is {\it walkable} if there is a directed
walk $W$ in $D$ from $s$ to $t$ such that the set of arcs traversed by $W$ (possibly repeatedly) is $S$.

Now consider how the information about $G$ and its turncost functions is represented in $\mathcal{M}(G)$.  A 
$k$-turn covering walk 
$W$ in $G$ that starts at a vertex $s$ and ends at a vertex $t$ is described by the information: \\
(1) a sequence of $k+2$ vertices: $s=x_{0}, x_{1},..., x_{k+1}=t$, and \\
(2) a sequence of $k+1$ subwalks $W_{0},...,W_{k}$ where for $i=0,...,k$, $W_{i}$ is a turncost-free walk from
$x_{i}$ to $x_{i+1}$, that has the property that every vertex of $G$ is visited on at least one of the subwalks.  

Let $\mathcal{D}(G)$ be the subdigraph of $\mathcal{M}(G)$ induced by the vertices of
$\mathcal{V}_{2}$.  A turncost-free walk in $G$ corresponds to a directed walk in $\mathcal{D}(G)$, 
and vice versa, by the definition of $\mathcal{A}_{5}$.

\medskip
\noindent
{\it Claim 1.}  $G$ admits a $k$-turn covering walk if and only if:
\begin{enumerate}
\item[(1)] there are $k+2$ vertices $x_{0},...,x_{k+1}$ of $\mathcal{V}_{1}$ in $\mathcal{M}(G)$, and
\item[(2)] $k+1$ purposeful sets of arcs $(S_{i}, s_{i}, t_{i})$ in $\mathcal{D}(G)$, $0 \leq i \leq k$, such that

\begin{itemize}
\item[(i)] $(S_{i},s_{i}, t_{i})$ is walkable in $\mathcal{D}(G)$ for $i=0,\ldots ,k$,
\item[(ii)] there is an arc in $\mathcal{A}'$ from $t_{i}$ to $x_{i+1} \in \mathcal{V}_{1}=V$, 
and from $x_{i}$ to $s_{i}$, for $i=0,\ldots ,k$, and
\item[(iii)] for every vertex $x \in \mathcal{V}_{1}=V$, there is some index $j$, $0 \leq j \leq k$,
and an arc $a=u \cdot v \in S_j$, such that there is an arc in $\mathcal{A}'$ in either direction,
between $x$ and $u$ or $v$. 
\end{itemize}
\end{enumerate}

In one direction, the claim is easy: given a $k$-turn covering walk $W$ in $G$, it is naturally factored into
$k+1$ turncost-free subwalks $W_{i}$, and each traversal of an edge of $G$ in a subwalk $W_{i}$ corresponds in
$\mathcal{M}(G)$ to a visit to a vertex of $\mathcal{V}_{2}$, thus the sequence of edge transversals of 
$W_i$ in $G$ corresponds 1:1 with a sequence $(y_{0},...,y_{m+1})$ of vertices of $\mathcal{V}_{2}$ in
$\mathcal{M}(G)$.  Because $W_i$ is turncost-free, by the definition of $\mathcal{A}_{5}$, there is an
arc in $\mathcal{D}(G)$ from $y_{i}$ to $y_{i+1}$ for $i=0,...,m$.  We take the set of arcs to be $S_i$,
$s_{i}=y_{0}$ and $t_{i}=y_{m+1}$, giving us (1) and (2) in a well-defined manner.  It is straightforward to
check that the conditions hold.  For example, the assumption that $W$ is a covering walk in $G$ yields the
last condition.

Conversely, suppose we have (1) and (2) in $\mathcal{M}(G)$.  
By the second condition, each $S_i$ is walkable.
By the definition of $\mathcal{A}_{5}$, a directed
walk for $S_i$ in $\mathcal{D}(G)$ corresponds to a turncost-free walk $W_i$ in $G$.
The third condition insures that the subwalks $W_i$ in $G$ can be sequenced into a $k$-turn walk $W$, where
the turns occur at the vertices $x_i$ by the first condition.  $W$ is covering in $G$ by the fourth condition,
yielding Claim 1.
\smallskip

\noindent
{\it Claim 2.} Consider a digraph $D=(V,A)$ equipped with distinguished vertices $s$ and $t$ (allowing $s=t$).
The property: ``{\it there exists a directed walk from $s$ to $t$ that traverses (allowing repetition) every arc in $A$}''
(that is, $(A,s,t)$ is walkable) is expressible in MSO logic.

We first argue that $(A,s,t)$ is walkable if and only if there is a directed path $P$ in $D$ from $s$ to $t$, such that every arc
$a \in A$ either is an arc of $P$, or belongs to a strongly connected subdigraph $D_{a}$ that includes a vertex of $P$.  
We then argue (in Appendix A) that this property is expressible in
MSO logic in a straightforward manner.

Given such a directed path $P=(s=x_{0},...,x_{m}=t)$ in $D$, we can describe a walk $W$ that traverses every
arc of $A$ as follows.  
By the {\it arcs of $P$} we refer to the set of arcs 
$$A[P]= \{ x_{0}x_{1}, x_{1}x_{2},..., x_{m-1}x_{m} \}$$
The walk has $m$ phases, one for each vertex $x_i$ of the path $P$.
Partition the arcs of $A-A[P]$ into $m$ classes $A_{0},...,A_{m}$ where for $i=1,...,m$ every arc $a=uv \in A_i$
belongs to a strongly connected subdigraph $D_{a}$ that includes the vertex $x_i$.  
Such a partition exists, by the supposed property of $P$.
There is a directed path
in $D_a$ from $x_i$ to $u$, and from $v$ to $x_i$, by the strong connectivity of $D_a$, and so there is a directed
cycle in $D_a$ that includes both $a$ and $x_i$.  Include this cycle in $W$, starting from $x_i$ and returning to $x_i$,
for each arc $a \in A_i$.  Increment $i$, take the arc from $x_{i}$ to $x_{i+1}$ and repeat this for
$i=0,...,m$.  

Now suppose that there is a directed walk $W$ in $D$ from $s$ to $t$ that traverses every arc in $A$.  
If there is a vertex $v$ that is visited more than once, then we can find a shorter walk $W'$ that, considered as
a sequence of arc transversals, is a subsequence of the sequence of arc transversals of $W$.  Therefore, 
by downward induction, there is a
directed path $P$ from $s$ to $t$, with no repeated internal vertex visits, that considered as a sequence of arc transversals, is a subsequence of the sequence of arc transversals of $W$.  But then, every arc $a$ traversed in the walk
$W$ (that is, every arc $a \in A$), that is not an arc of $P$, must belong to a subwalk $W'$ of $W$ that begins 
and ends at a vertex of $P$.  The vertices visited by $W'$ therefore induce a strongly connected subdigraph containing
a vertex of $P$.

The second part of the proof of Claim 2 is to argue the property we have identified is expressible in MSO logic. 
This is fairly routine for MSO (see Appendix A).
Based on Claims 1 and 2, 
the remainder of the proof of Theorem 3 is also straightforward (Appendix A). \hfill $\Box$
\end{proof}


\section{Discrete Milling is Hard for Bounded Pathwidth}
\label{sec:discrete-milling-hardness}

In this section, we see that the maximum degree restriction implicit in the parameterization for our positive
result in the last section is one of the keys to tractability for this problem. 
Recall that in the {\sc Discrete Milling} problem the turncost functions are assumed pairing and symmetric.
This is a significant assumption, but the outcome is still negative, and the following result very much
strengthens, in the parameterized setting, the NP-completeness result of Arkin, et al. \cite{ABD05}.

\begin{theorem} \label{th:paired}
{\sc Discrete Milling} is hard for $W[1]$,  when
parameterized by $(k,p)$, where $k$ is the number of turns and $p$
is a bound on pathwidth.
\end{theorem}
\begin{proof}
The fpt-reduction is from {\sc Multicolor Clique}, using an edge
representation strategy, such as described, for example, in
\cite{FFL07,Sze08
}. 
Suppose $G=(V,E)$ has $V$ partitioned into  color  classes
$C_i$, $i=1,...,r$.  The {\sc Multicolor Clique} problem asks
whether $G$ contains a $r$-clique consisting of one vertex from each
color class $C_i$.  We can assume that each color class of $G$ has
the same size $n$ \cite{FFL07}.  The color-class partition of $V$
induces a partition of $E$ into ${{r}\choose{2}}$ classes
$E_{\{i,j\}}$, for $1 \leq i \not= j \leq r$:
$$ E_{\{i,j\}} = \{ e \in E: \exists u \in C_i \mbox{~and~} \exists v \in C_j
\mbox{~with $e$ incident on $u$ and $v$~} \} \, .$$ We can also
assume that all these edge-partition  classes $E_{\{i,j\}}$ have
the same size $m$.
We index  the  vertices and edges of $G$ as follows:
\vspace{-0.1cm}
\begin{eqnarray*}
C_i & = & \{ v(i,q): 1 \leq q \leq n \} ~~~ \mbox{for $i=1,...,r$} \\
E_{\{i,j\}} & = & \{ e(\{i,j\},l): 1 \leq l \leq m \} ~~~ \mbox{for $1 \leq i \not=
j \leq r $} .
\end{eqnarray*}

To refer to the incidence structure of $G$,  we define  functions
$\pi_{\{i,j\}}^{i}(l)$ and $\pi_{\{i,j\}}^{j}(l)$ as follows:
\vspace{-0.1cm}
\begin{eqnarray*}
\pi_{\{i,j\}}^{i}(l) & = q:~ \mbox{ the edge } e(\{i,j\},l) \mbox{~is incident on~} v(i,q) \\
\pi_{\{i,j\}}^{j}(l) & = q~ \mbox{ the edge } e(\{i,j\},l) \mbox{~is incident on~}
v(j,q) \, ,
\end{eqnarray*}
so the edge $e(\{i,j\},l)$ is incident to $v(i,\pi_{\{i,j\}}^{i}(l))$ and
$v(j,\pi_{\{i,j\}}^{j}(l))$.

We describe the construction of a graph $G'$, together with the
sets $S_v$ of turn-free pairs of  edges for the vertices $v$ of
$G'$.  We first describe the vertices of $G'$, and
then specify a set of paths on these vertices.  The edge set of the
multi-graph $G'$ is the (abstract) disjoint union of the sets of
edges of these abstractly-defined paths, and it is understood that
each path is turn-free, so that (for the most part), the sets
$S_v$ of turn-free pairs of $v$-incident edges for the vertices
$v$ of $G'$ are implicit in these {\it generating paths} of $G'$.


The vertex set $V'$ for $G'$ is the union of the sets
$V_{0} \cup V_{1} \cup V_{2} \cup V_{3} \cup V_{4}$,
\begin{eqnarray*}
V_{0} & = & \{ \sigma, \tau \} \\
V_{1} & = & \{ t[i,j]: 1 \leq i \not=j \leq r \} \\
V_{2} & = & \{ s[i,j]: 1 \leq i \not= j \leq r \} \\
V_{3} & = & \{ c[i,j,u]: 1 \leq i \not= j \leq r,~1 \leq u \leq n \} \\
V_{4} & = & \{ p[i,j,l]: 1 \leq i \not= j \leq r,~ 1 \leq l \leq m \} .
\end{eqnarray*}

Thus $|V_{1}| = |V_{2}| = 2{{r}\choose{2}}$, $|V_{3}| = 2n{{r}\choose{2}}$
and $|V_{4}| = 2m{{r}\choose{2}}$.

The edge set of $G'$ is (implicitly) described by a
generating set of paths ${\cal P}$ (two paths for each edge of $G$),
together with a few more edges:
$$ {\cal P} = \{ P[i,j,e(\{i,j\},l)]: 1 \leq i \not= j \leq r,~ 1 \leq l \leq m \,
, \}  \, ,  $$
where the path $P[i,j,e(\{i,j\},l)]$
(1) starts at the vertex $p[i,j,l]$;
(2) next visits $s[i,j]$;
(3) then visits the vertices $c[i,j,u]$, except
for $u= \pi_{\{i,j\}}^{i}(l)$ (the
{\it exceptional vertex}
of this block), in {\it consecutive order}, meaning that the vertices are
visited by increasing index $u$,
modified by skipping the exceptional vertex;
(4) then visits the vertex $c[i,j^{*},\pi_{\{i,j\}}^{i}(l)]$, where $j^{*}$ is
defined to be $j+1$,
unless $j+1=i$, when $j^{*}=j+2$,
or $j=r$ and $i \not= 1$, when $j^{*}=1$,
or $j=r$ and $i=1$, when $j^{*} = 2$; and then
(5) ends at the vertex $t[i,j]$.

Intuitively, there are two paths in ${\cal P}$  corresponding to
each edge of $G$.  If we fix $i$ and consider that there are $r-1$
blocks of vertices (each block consisting of $n$ vertices,
corresponding to the vertices of $C_i$), then what a path
$P[i,j,e(\{i,j\},l)]$ (corresponding to the $l^{th}$ edge of
$E_{\{i,j\}}$) does is ``hit'' every vertex of its ``own''
$\{i,j\}^{th}$ block, except the vertex
$c[i,j^{*},\pi_{\{i,j\}}^{i}(l)]$ of the block corresponding to
the vertex of $C_i$ to which the indexing edge of $G$ is incident,
and in the ``next block'' in a circular ordering of the $r-1$
blocks established by the definition of $j^{*}$, does the
complementary thing: in this ``next block'' it hits {\it only} the
vertex corresponding to the vertex of $C_i$ to which the indexing
edge is incident in $G$, and then ends at $t[i,j]$.

At this stage of the construction, the edges of $G'$  are
partitioned into (turn-free) paths that run between vertices of
$V_1$ and vertices of $V_4$, where the latter have degree 1 (so
far) and the vertices of $V_1$ have degree $m$ (so far). We
complete the construction of $G'$ by adding a  few more edges, 
specifying a few more turn-free pairs as we do so.

(A) Add edges between the pairs of vertices $p[i,j,l]$ and
$p[j,i,l]$ for all $1 \leq i \not= j \leq r$ and $1 \leq l \leq
m$. After these edges are added, we have reached a stage  where
all vertices in $V_4$ have degree 2 (and they will have degree 2
in $G'$).  For each vertex of $V_4$ we make the pair of incident
edges a turn-free pair.

Note that for any instance of the {\sc Discrete Milling},
the edge set is naturally and uniquely partitioned into maximal
turn-free paths. At this stage of the construction, these paths
all run between $t[i,j]$ and $t[j,i]$ for $1 \leq i < j \leq r$.

(B) Add some edges between the vertices of $V_0 \cup V_1$.   Let
$\leq_{lex}$ denote the lexicographic order on the set (of pairs
of indices) ${\cal I} = \{[i,j]: 1 \leq i<j \leq r\}$.  Let
$[i,j]^{*}$ denote the immediate successor of $[i,j]$ in the
ordering of ${\cal I}$ by $\leq_{lex}$.  For $[i,j] \in {\cal I}$,
let $rev[i,j] = [j,i]$.  We add the edges (using the notation $u
\cdot v$ to denote the creation of an edge between
$u$ and $v$):

\noindent
(B.1) $t[rev[i,j]]\cdot t[[i,j]^{*}]$ for $1 \leq i < j \leq r$ and $[i,j] \not= [r,r-1]$,\\
(B.2) $\sigma \cdot t[1,2]$, and\\
(B.3) $t[r,r-1] \cdot \tau \, .$

We do not specify any further turn-free pairs of vertex
co-incident  edges beyond those specified in (A) or implicit by
being internal to the generating paths ${\cal P}$ of $G'$. That
completes the description of $G'$.

To complete the proof, we need to show that: (1) the graph $G'$
will admit a $k$-turn covering walk, where $k= 2
{{r}\choose{2}}$, if and only if $G$ has a multicolor $r$-clique;
and (2) $G'$ has path-width at most $6{{r}\choose{2}} + 4$. (See
Appendix A)

\end{proof}


\section{Open Problems}


We have studied the parameterized complexity of (several versions of) the abstract milling 
problem with turn costs and gave an initial classification with respect to several parameterizations. 
Our FPT results are impractical, but can they be improved?  
In particular, it would be interesting to know if {\sc Abstract Milling} parameterized by $(k,t,d)$ admits a
polynomial kernel.
Our negative result provides one of the very few natural examples of a parameterized graph problem
that is $W[1]$-hard, parameterized by pathwidth.  
Another notable open question is whether {\sc Discrete Milling} parameterized by $(k,d)$, 
is FPT or W[1]-hard.

\vspace{-0.1cm}
\bibliographystyle{plain}


\newpage
\section*{Appendix A}

\bigskip
\begin{proofof} \textit{\textbf{Lemma~\ref{lem:msol-grid}:}}

\noindent
Testing whether the graph induced by a set $S$ of vertices is a monochromatic
path in the grid graph $G$ between two distinct vertices $u$ and $v$ of $S$ can be done by
the following formula:

\noindent\small
\begin{eqnarray}
\LinPath(u,v,S)=~~~~~~~~~~~~~~~~~~~~~~~~~~~~~~~~~~~~~~~~~~~~~~~~~~~~~~~~~~~~~~~~~~~~~~~~\\
\Big\{\exists x:\big\langle Sx\wedge x\neq u\wedge Eux\big\rangle 
\wedge 
\big\langle \forall x': (Sx'\wedge x\neq x'\wedge x\neq u ) \rightarrow \neg Eux'\big\rangle\Big\}\\
\bigwedge~~
\Big\{\exists y:\big\langle Sy\wedge y\neq v\wedge Evy\big\rangle 
\wedge 
\big\langle \forall y': (Sy'\wedge y\neq y'\wedge y'\neq v ) \rightarrow \neg Evy'\big\rangle\Big\}\\
\bigwedge~~~
\Big\{\big\langle \exists e=yx: (Sx\wedge Sy\wedge E_he)\big\rangle 
\rightarrow 
\big\langle \forall e'=x'y': (Sx'\wedge Sy') \rightarrow E_he'\big\rangle \Big\}\\
\bigwedge~~~
\Big\{\big\langle \exists e=yx: (Sx\wedge Sy\wedge E_ve)\big\rangle 
\rightarrow 
\big\langle \forall e'=x'y': (Sx'\wedge Sy') \rightarrow E_ve'\big\rangle \Big\}\\
\bigwedge~~~
\Big\{ \forall S_1,S_2: \big\langle\forall x: Sx\rightarrow  [(S_1x\leftrightarrow \neg S_2x)\wedge (S_1x\vee S_2x)]\big\rangle~
\rightarrow~~~~~~~~~~~~~\\
\big\langle\exists x_1, \exists x_2: (S_1x_1\wedge S_2x_2 \wedge Ex_1x_2) \big\rangle \Big\} .
\end{eqnarray}

\normalsize
Lines (2, (3) assert that $u$ and $v$ have only one neighbor in $S$. 
Lines (4), (5) respectively check whether all the edges of $G$ with both endpoints in $S$ are horizontal or vertical (actually only one of the two lines is needed). Note that when no edge in $G$ has both endpoints in $S$, both lines (3) and (4) become true, however this is taken care of by the next implication which checks connectivity.  
The implication (6)$\rightarrow$ (7) guarantees that for any bipartition $(S_1,S_2)$ of $S$
there is an edge from $S_1$
 to $S_2$, i.e. $S$ induces a connected subgraph. Together with (2), (3), (4) and (5) this implies that
the graph induced by $S$ is a connected subgraph of $G$ with either vertical or horizontal edges, where both $u$ and $v$ have degree one. 
Since $G$ is a grid graph, each vertex has at most two horizontal and at most two vertical incident edges. 
Hence, the graph induced by $S$ is a path with extremities $u$ and $v$.

Finally, a $k$-covering walk exists if and only if there are $k+1$
subsets of vertices covering the vertex set of the input graph
such that each induces a monochromatic path and such that these linear
paths are joined end to end:

\small
\noindent
\begin{eqnarray}
\Phi=\exists v_0,\dots ,v_{k+1}, \exists S_0,\dots ,S_k : ~~~~~~~~~~~~~~~~~~~~~~~~~~~~~~~~~~~~~~~~~~~~~~~~~~~~~~~~~~~~~~~~\\
\big(\underset{0\leqslant i \leqslant k+1}{\bigwedge} Vv_i \big) 
\wedge
\big(\forall v : Vv \leftrightarrow \underset{0\leqslant i\leqslant k}{\bigvee} S_iv \big)
\wedge 
\big(\underset{0\leqslant i\leqslant k}{\bigwedge} S_{i}v_i \big)~~~~~~~~~~~~~~~~~~~~~\\
\wedge 
\big(\underset{0\leqslant i\leqslant k}{\bigwedge} S_{i}v_{i+1} \big)~~~~~~~~~~~~~~~~~~~\\
\wedge  
\big(\underset{0\leqslant i \leqslant k}{\bigwedge}\LinPath(v_i,v_{i+1},S_i) \big) .
\end{eqnarray}
\end{proofof}

\bigskip
\begin{proofof} \textit{\textbf{Theorem 2}:} (MSO)

Let $D=(V,A)$ be a digraph, and $s,t$ vertices of $D$ (allowing $s=t$).  

The first subtask (to finish the proof of Claim 2) is to describe an MSO predicate
that expresses
that there is a directed path $P$ in $D$ from $s$ to $t$, quantified on the sets of vertices and arcs
that form the path.

$$dipath(s,t) =  \exists U (\subseteq V) \exists B (\subseteq A):  ... $$

Where the remainder of the predicate expresses that in the subdigraph $D'=(U,B)$:

$\bullet$ $s$ has outdegree 1 and indegree 0

$\bullet$ $t$ has indegree 1 and outdegree 0

$\bullet$ every vertex of $U$ not $s$ or $t$ has indegree 1 and outdegree 1

$\bullet$ for every partition of $U$ into $U_1$ and $U_2$ such that $s \in U_1$ and $t \in U_2$,
there is a vertex $u \in U_{1}$ and a vertex $v \in U_2$ with an arc in $B$ from $u$ to $v$

The formalization is completed in a very similar manner to the details of the proof of Lemma 2 
(re: Linpath) above in this
Appendix.

Being able to express that there is a directed path from $s$ to $t$ leads easily to an MSO
predicate for strong connectivity of a subdigraph described by a set of vertices and a set of arcs.

An MSO predicate for {\it walkability} of a set of arcs $A$ relative to $s$ and $t$ is easily 
(but somewhat tediously) constructed
on the basis of the structural characterization of {\it Claim 2}, using the predicates for the
existence of an $s$-$t$ path, and for strongly connected subdigraphs.

An MSO formula to complete the proof of Theorem 3 is then trivial to construct by writing out
{\it Claim 1} in the formalism.

\end{proofof}

\bigskip
\begin{proofof} \textit{\textbf{Theorem~\ref{th:paired}:}} (Correctness)

\noindent
Figure~\ref{fig:W2} shows how the ``coherence'' gadgets for the reduction from {\sc Multicolor Clique} work
in our construction.

\begin{figure}[htb!]
\centering
\resizebox{!}{2.2in}
{\includegraphics{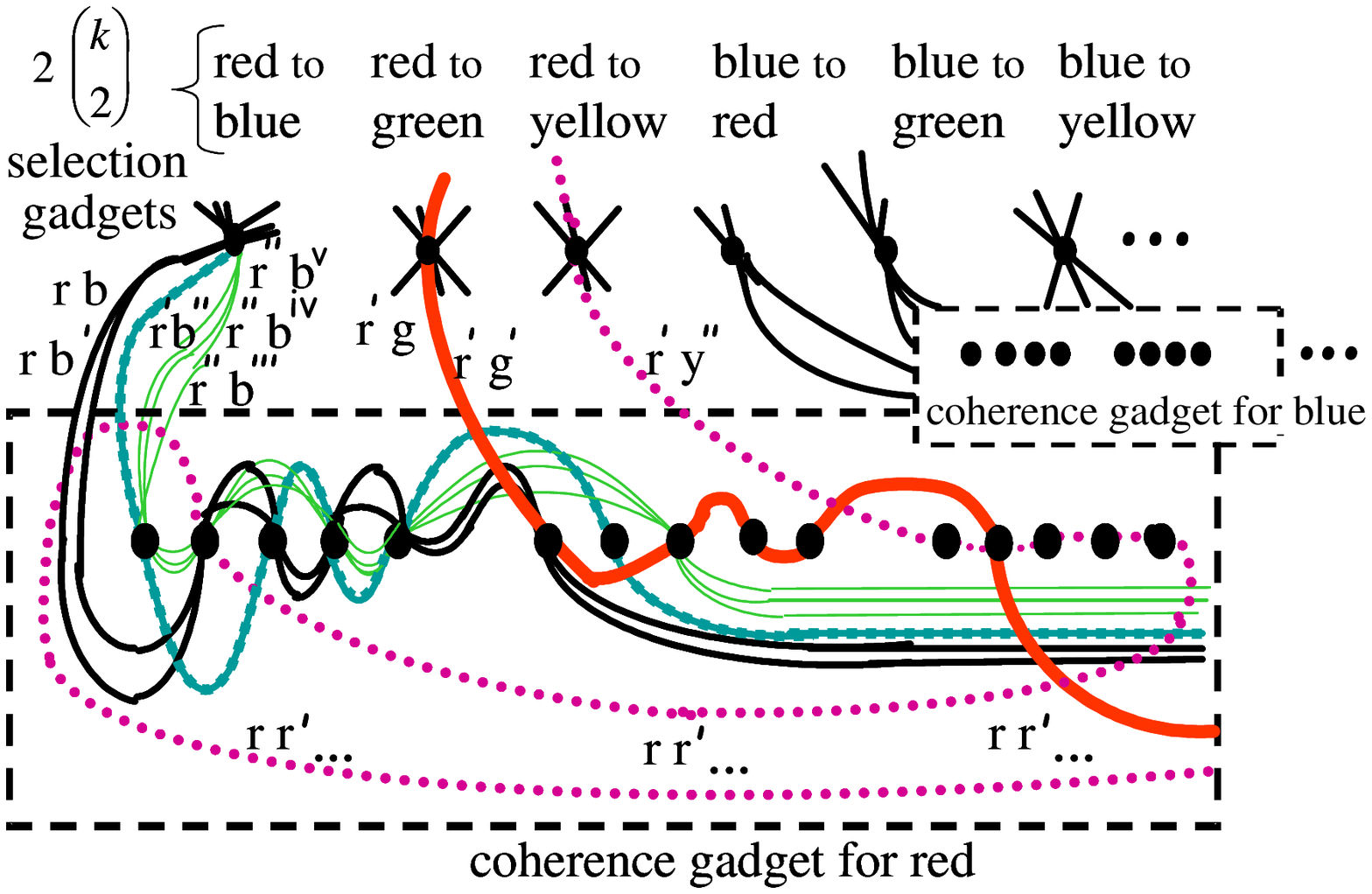}}
\caption{The choice and coherence gadgets for the reduction from {\sc Multicolor Clique}.
For this example,  $r =4$ and the vertex color classes $C_i$ are called
{\em red}, {\em blue}, etc.}
\label{fig:W2}
\end{figure}

Suppose $G'$ has a $k$-turn covering walk ${\cal W}$ that visits
all the vertices of $G'$.  (Recall that $k = 2{{r}\choose{2}}$.)
Note that every visit to a vertex of $V_{0} \cup V_{1}$ entails a
turn.  Since $|V_0 \cup V_1 | = k + 2$, this implies that the walk
${\cal W}$ must begin and end at vertices of $V_0 \cup V_1$, and
must visit each of the vertices in this set, internal to ${\cal
W}$, exactly once. We can further conclude that we can view ${\cal
W}$ as necessarily beginning at the vertex $\sigma$ of $G'$ and
ending at the vertex $\tau$, as otherwise, ${\cal W}$ would have
to visit $t[1,2]$ or $t[r,r-1]$ more than once, in order to visit
$\sigma$ and $\tau$.

Since the turns of the solution walk ${\cal W}$ occur (in each case exactly once) at the vertices of $V_1$,
we may conclude that ${\cal W}$ must consist of the edges added in (B), together with exactly ${{r}\choose{2}}$ maximal turn-free paths between $t[i,j]$ and $t[j,i]$ for $1 \leq i < j \leq r$.  By the indexing of the paths in
${\cal P}$ in the construction of $G'$, ${\cal W}$ therefore corresponds to a set of ${{r}\choose{2}}$ edges of $G$, where for each $i$, exactly $r-1$ of these edges are incident on a vertex of $C_i$.  We must argue that, for a given $i$,
these $r-1$ edges of $G$ are incident on {\it the same} vertex of $C_i$.  (This is the job of what is sometimes
called the {\it coherence gadget} for a reduction from {\sc Multicolored Clique} using the edge-representation
strategy --- which we are employing here.)  It is easy to check that the
circularly interlocking path structure (for fixed $i$) of the paths in ${\cal P}$, together with the fact that the solution walk ${\cal W}$ visits all the vertices of
$V_3$ (in particular, for the subset of $V_3$ formed by fixing $i$), insures that this is the case.

The converse direction is essentially trivial.

We next argue that the pathwidth of $G'$ is bounded by a function of $r$.  Let $G''$ be the subgraph of $G'$
formed by deleting from $G'$ the vertices of
$$V_0 \cup V_1 \cup V_2 \cup \{c[i,j,n]: 1 \leq i \not= j \leq r  \} \, . $$
Note that we are deleting only the {\it last} vertices of the $2{{r}\choose{2}}$ blocks of $n$ vertices
(in each block) of the coherence gadgets of the construction (see Figure~\ref{fig:W2}).  The total number of vertices
deleted is $6{{r}\choose{2}}+2$.  It is easy to check that $G''$ has pathwidth at most 2, and therefore
$G'$ has pathwidth at most $6{{r}\choose{2}} + 4$.
\end{proofof}

\section*{Appendix B}

\subsection*{Parameterized complexity}

A problem is \emph{fixed-parameter tractable} with respect to a
parameter $k$ if it can be solved in $O(f(k)\cdot n^{O(1)})$ time,
where $n$ is the size of the input and $f$ is a computable
function depending only on $k$; such an algorithm is (informally)
said to run in fpt-time. The class of all fixed-parameter
tractable problems is denoted by FPT. For establishing
fixed-parameter intractability, a hierarchy of classes, the
W-hierarchy, has been introduced: a parameterized problem that is
hard for some level of W is not in FPT under standard complexity
theoretic assumptions about the difficulty of quantified forms of the
{\sc Halting Problem} for nondeterministic Turing machines \cite{DF99}.
Hardness is sought via fpt-reductions: an
\emph{fpt-reduction} is an fpt-time Turing reduction from a
problem $\Pi$, parameterized with $k$, to a problem $\Pi'$,
parameterized with $k'$, such that $k'\leq g(k)$ for some
computable function $g$.

\subsection*{Tree decompositions and graph minors}

\begin{definition}
A \emph{tree decomposition} of a graph $G=(V,E)$ is a pair $(T,\mathcal{X})$,  where
$T$ is a tree and $\mathcal{X}$ is a family of subsets (bags) of $V$ such that:
1) for any vertex $x\in V$, there exists $X\in\mathcal{X}$ such that $x\in X$;
2) for any edge $e=(x,y)\in E$, there exists $X\in\mathcal{X}$ such that
$\{x,y\}\subseteq X$; and
3) for any vertex $x\in V$, the set of bags containing $x$ induces a subtree $T_x$ of $T$.
The \emph{tree-width} of a graph is:
$$tw(G)=\min_{(T,\mathcal{X})} width(T,\mathcal{X})=\min_{(T,\mathcal{X})}
\max_{X\in\mathcal{X}} (|X|-1) \, .$$
\end{definition}

The notions of path decomposition
and path-width are defined similarly by changing the tree $T$ in
the above definition to a path $P$. The path-width of a graph is always larger than 
or equal to its tree-width. 
Computing the tree-width of a graph $G$ is NP-hard, but deciding
whether the tree-width of $G$ is at most $k$ (and computing a tree-decomposition of 
width at most $k$ in the positive case) is fixed-parameter tractable (c.f. \cite{FG06}). 

For an edge $e=xy$ of a graph $G$, contracting $e$ results in
replacing the vertices $x$ and $y$, whose neighborhoods are
denoted by $N(x)$ and $N(y)$ respectively, by a new vertex $z$
whose neighborhood is $N(x)\cup N(y)$, excluding $x$ and $y$. A
graph $H$ is a \emph{minor} of a graph $G$ if $H$ can be obtained
from $G$ by a series of edge or vertex removals and edge
contractions.

\newpage
\subsection*{Monadic second order logic}

MSO logic for annotated graphs gives us:\\

-- variables denoting individual vertices ($s,t,u,v,...$, possibly with subscripts) and variables denoting
sets of vertices ($S,T,U,V,...$, possibly with subscripts).  We will use the notation $Tt$ to denote that the vertex
$t$ belongs to the set of vertices $T$, and similarly in general. \\
-- variables denoting individual arcs and edges ($a,b,...,e,f,...$) and sets of arcs and edges 
($A,B,...,E,F,...$ )\\
-- logical quantification over these variables, logical connectives \\
($\land, \lor, \implies, \iff, \neg, ...$) and equality ($u=v, E=F,...$) \\
-- predicates ($inc(e,u), adj(u,v), in(a,u), out(a,v) ...$) indicating incidence of edges and vertices, and arcs
entering or leaving vertices \\
-- predicates such as $A_{r}a$ ($a$ is an arc of type $r$) and shorthands such as
$\exists a = uv$ (there exists an arc $a$ from $u$ to $v$).

\begin{theorem} [Courcelle \cite{Cou90}]
Any property of annotated graphs that can be expressed in MSO, is linear-time, finite-state recognizable
for annotated graphs where the treewidth of the underlying graphs is bounded.
\end{theorem}

\end{document}